%% file: main.tex
\documentclass[sigconf]{acmart}

\usepackage[]{subcaption}
\usepackage{cleveref}
\usepackage{caption}
\usepackage{color}
\usepackage{colortbl}
\usepackage{xcolor}
\usepackage{xspace}
\usepackage{comment}

\usepackage{subcaption}
\usepackage{array}
\usepackage{multirow}
\hypersetup{
    colorlinks=true,
    linkcolor=red,
    filecolor=red,      
    urlcolor=red,
    citecolor=red
    }
\usepackage{longtable}
\usepackage{tabularx}
\usepackage{url}
\usepackage[flushleft]{threeparttable}
\usepackage[online]{threeparttablex}
\usepackage{enumitem}
\AtBeginDocument{%
  \providecommand\BibTeX{{%
    \normalfont B\kern-0.5em{\scshape i\kern-0.25em b}\kern-0.8em\TeX}}}

\setlist{itemsep=0pt,parsep=0pt}
\usepackage[export]{adjustbox}
\usepackage[english]{babel}
\newcommand\Mu{\mathrm{M}}

\acmDOI{}
\acmISBN{}
\acmPrice{}
\renewcommand\footnotetextcopyrightpermission[1]{} 
\setcopyright{none}

\begin{document}
\title{Semantic Caching for Improving Web Affordability}
\author{
Hafsa Akbar, Danish Athar, Muhammad Ayain Fida Rana, Chaudhary Hammad Javed, \\ Zartash Afzal Uzmi, Ihsan Ayyub Qazi,  Zafar Ayyub Qazi \vspace{0.5em}
}

\affiliation{%
  \institution{Lahore University of Management Sciences}
  \country{Pakistan}
}

\renewcommand{\shortauthors}{Hafsa Akbar et al.}

\raggedbottom
\begin{abstract}
\input{abstract.tex}

\end{abstract}

\maketitle

\input{intro}

\input{methodology.tex}
\input{potential.tex}

\input{models.tex}
\input{analysis.tex}

\input{discussion}
\input{related_work}

\input{conclusion.tex}

\bibliographystyle{ACM-Reference-Format}
\bibliography{main}

\input{appendix.tex}

\end{document}

%% file: abstract.tex
The rapid growth of web content has led to increasingly large webpages, posing significant challenges for Internet affordability, especially in developing countries where data costs remain prohibitively high. We propose semantic caching using Large Language Models (LLMs) to improve web affordability by enabling reuse of semantically similar images within webpages. Analyzing 50 leading news and media websites, encompassing 4,264 images and over 40,000 image pairs, we demonstrate potential for significant data transfer reduction, with some website categories showing up to 37\% of images as replaceable. Our proof-of-concept architecture shows users can achieve approximately 10\% greater byte savings compared to exact caching. We evaluate both commercial and open-source multi-modal LLMs for assessing semantic replaceability. GPT-4o performs best with a low Normalized Root Mean Square Error of 0.1735 and a weighted F1 score of 0.8374, while the open-source LLaMA 3.1 model shows comparable performance, highlighting its viability for large-scale applications. This approach offers benefits for both users and website operators, substantially reducing data transmission. We discuss ethical concerns and practical challenges, including semantic preservation, user-driven cache configuration, privacy concerns, and potential resistance from website operators. 

%% file: intro.tex
\section{Introduction}
\label{sec:intro}

\begin{figure}
  \centering
  \includegraphics[width=0.45\textwidth]{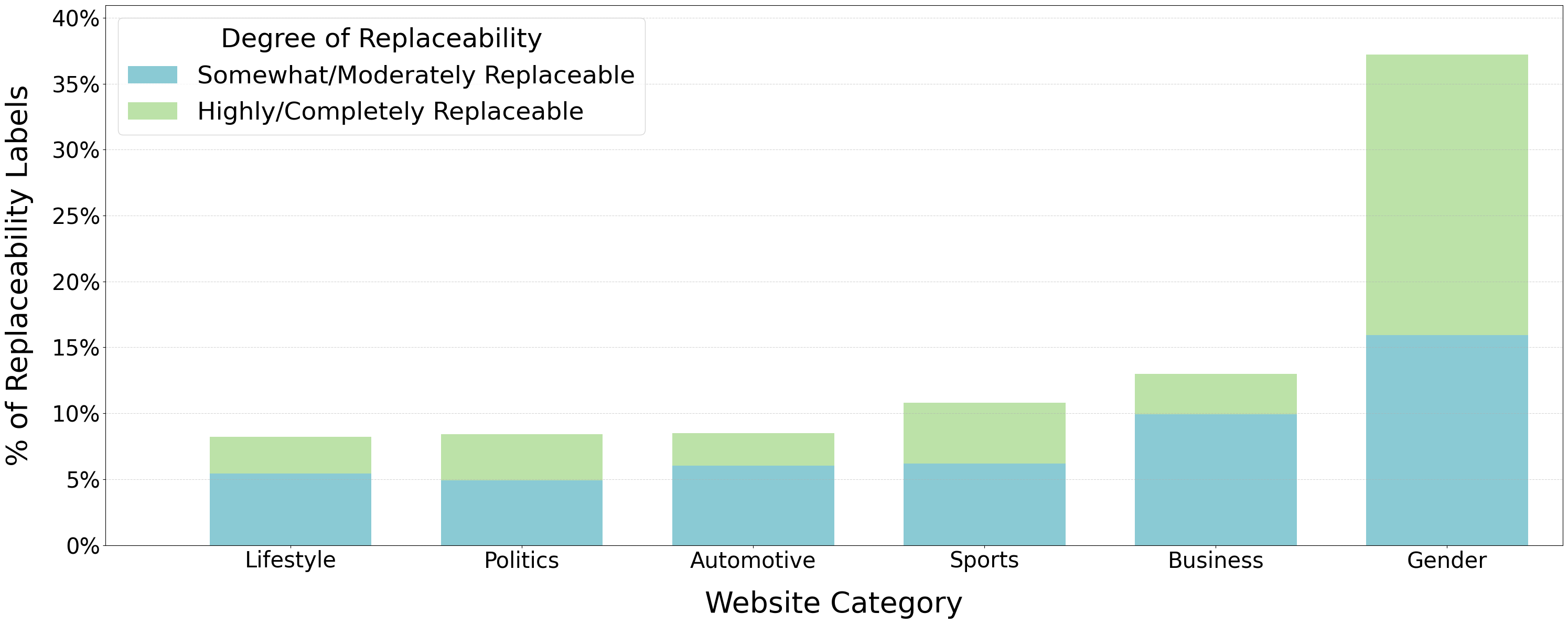}
  \vspace{-0.15in}
  \caption{\small{Image replaceability potential across website categories. Based on a user study of top 50 news websites, showing percentages of somewhat/moderately and highly/completely replaceable images per category.}}
  \label{fig:category_savings}
\end{figure}

The rapid growth of the Web has led to increasingly large webpages, posing significant challenges for Internet affordability, especially in developing countries. Over the past decade, the median webpage size has approximately doubled, driven largely by the inclusion of more and higher-quality images, which typically account for 25\% to 60\% of the total page size~\cite{httparchive2024}. This surge exacerbates the digital divide, making Internet access prohibitively expensive for many users and, consequently, preventing them from accessing the vast opportunities the Internet has to offer. 

The issue is particularly acute in developing countries, where mobile devices are the primary mode of Internet access and data costs are disproportionately high relative to average income levels~\cite{mobile_statista}. A World Bank study across 11 emerging countries revealed that nearly half of the respondents struggled to pay for mobile data usage, with 42\% imposing self-restrictions, severely limiting their ability to fully engage with online resources and services~\cite{worldbankdd}. This is further underscored by the fact that, as of 2021, 95 countries failed to meet the UN Broadband Commission's affordability benchmark, which targets broadband costs at 2\% or less of the monthly Gross National Income (GNI) per capita~\cite{itu, target_broadband}.\footnote{This is for a 2\,GB data-only mobile broadband plan.}

To help address the growing challenge of increasing webpage size and improve Internet affordability, we explore the potential for \textit{semantic caching} on the Web. Specifically, we focus on the semantic caching of images, whereby clients can reuse similar or replaceable images already stored in their local cache instead of downloading new ones. To assess the potential of semantic caching on the Web, we conducted  an analysis of 50 leading news and media websites, encompassing over 40,000 image pairs. Our study involved human evaluators who assessed the replaceability of images from recent articles within the same website and category.\footnote{Human annotators assigned replaceability scores to image pairs, considering factors like thematic consistency, relevance, and potential for context loss.} The results reveal significant opportunities for reducing data transfer in web content delivery through image reuse.

Figure~\ref{fig:category_savings} summarizes our findings across various content categories, such as gender, sports, business, and politics.\footnote{These categories were sampled as part of our dataset and grouped based on the mapping in Appendix \ref{appendix:category-map}.} A considerable percentage of images demonstrate a reasonable degree of replaceability. We define `feasible image replacement' as any instance where the replaceability meets or exceeds the lowest threshold described as `somewhat replaceable'. Notably, gender-related content shows the highest potential, with up to 37\% of feasible replacement (with already cached images from same website and category\footnote{Our dataset comprised of images sourced from the 10 latest articles within a category for top 50 news and media websites}) of which most are highly/completely replaceable. This is due to abundant thematic similarity within the Gender category, where images often depict symbols of inclusivity, such as pride flags or LGBTQIA+ icons, or feature recurring political figures discussing gender rights leading to highly similar images across multiple articles.

Furthermore, we conduct a series of simulations where clients make a varying number of image requests across a diverse set of categories. For each article accessed, we estimate reduction of $6.4\%$ of the total page weight, with users achieving up to $\approx 10\%$  greater data savings on average compared to exact caching (\S\ref{sec:proof_of_concept}), assuming image replaceability is applied at the lowest level. These findings strongly suggest that implementing semantic caching could significantly reduce bandwidth requirements, especially for frequently updated content like news articles. 

To enable semantic caching on the Web, we propose a novel approach that leverages Large Language Models (LLMs). Recent advancements in LLMs have unlocked new possibilities in semantic analysis, offering robust capabilities for understanding and comparing visual content within context. In this study, we explore the application of state-of-the-art (SOTA) multi-modal LLMs—including GPT-4o, Gemini-1.5 Pro, and Claude 3.5 Sonnet—as well as the vision-language model LLaVA-NeXT combined with LLaMA 3.1, to measure the replaceability of images. Our research employs two distinct approaches:

\begin{enumerate}[leftmargin=*,topsep=0pt]
\setlength{\itemsep}{0pt}
\setlength{\parskip}{0pt} 
\item \textbf{Direct Image-to-Output Pipeline} utilizing proprietary models for immediate output based on image comparisons
\item \textbf{Two-Step Image Description-to-Output Pipeline} employing open-source models for a cost-effective solution suitable for large-scale applications.
\end{enumerate}

In our evaluation, GPT-4o achieved a balanced performance with a low Normalized Root Mean Square Error (NRMSE) of 0.1735 and a weighted F1 score of 0.8374. Notably, GPT-4o significantly outperformed other models with a Weighted Cohen's Kappa score of 0.62, indicating substantial agreement with human judgments. This model effectively balanced precision and recall across all classes, making it reliable for determining image replaceability.

Moreover, our experiments showed that the open-source LLaMA 3.1 model, despite being more cost-effective, performed comparably to commercial state-of-the-art models, highlighting the viability of open-source alternatives for large-scale applications. LLaMA outperformed all models in correctly classifying minority classes (all feasible replacements) and achieved SOTA performance in weighted precision. 

While our approach demonstrates significant potential for reducing data transfer and improving Internet affordability, implementing semantic web caching presents several challenges. Determining context loss when swapping images is inherently subjective, necessitating a sophisticated prompt design process that incorporates best practices in prompt engineering to ensure accurate evaluation. Secondly, the effectiveness of our caching system directly depends on the cache size; a greater number of relevant cached images provides more opportunity for replaceability. Additionally, the potential for image reuse differs across content categories, necessitating flexible caching strategies that can adapt accordingly. 

Our proposed solution offers mutual benefits for both users and website operators. By enabling clients to reuse cached images, operators can significantly reduce data transmitted from their servers, leading to cost savings and potentially improved load times, which enhance user engagement and satisfaction. Moreover, making content more accessible aligns with corporate social responsibility initiatives aimed at bridging the digital divide. To address concerns about content control, website operators can maintain authority over their content by flagging non-cacheable images using specific HTTP headers or metadata tags, and by setting caching policies that define how content is cached and reused, providing flexibility and control over content distribution.

Overall, our contributions can be summarised as 
follows:
\begin{itemize}[leftmargin=*,topsep=0pt]
\setlength{\itemsep}{0pt}
\setlength{\parskip}{0pt} 
\item We compiled and analyzed a comprehensive dataset from the top 50 news and media websites, encompassing 4,264 images and over 40,000 unique image pairs. This dataset was annotated by human evaluators to assess the potential for semantic caching. We have shared our dataset and code, it can be found here: \url{https://anonymous.4open.science/r/semantic-caching-www25-DCB1/} 
\item We developed and implemented a proof-of-concept architecture for semantic web caching, empirically demonstrating significant reductions in data transfer through byte savings measurements.
\item We conducted a comparative assessment of various SOTA LLMs for evaluating image replaceability, providing detailed insights into their effectiveness, limitations, and potential applications.
\item We introduced a novel, open-source approach using LLaVA-NeXT and LLaMA 3.1, offering a scalable and cost-effective solution for large-scale semantic caching applications. We also include a cost analysis of the two pipelines.
\item We critically examined the ethical implications and practical challenges of deploying semantic web caching systems, discussing key concerns such as privacy and content control and potential mitigation strategies.
\end{itemize}

The remainder of this paper is structured as follows: We detail our methodology (\S\ref{sec:methodology}), explore the potential of semantic web caching (\S\ref{sec:potential}), and analyze the role of LLMs (\S\ref{sec:llms}). We then present our implementation and results (\S\ref{sec:implementation}), discuss ethical concerns, deployment challenges, and limitations (\S\ref{sec:discussion}), review related work (\S\ref{sec:related_work}), and conclude (\S\ref{sec:concl}). 

%% file: methodology.tex
\section{Methodology}
\label{sec:methodology}
In this section, we outline the steps taken to collect the data and evaluate the models used.
\subsection{Data Collection}
We gathered data from the top 50 global news and media websites, ranked by traffic share for August 2024. News and media websites were specifically chosen for several key reasons: their high volume and frequency of image updates, clear topical categorization, contextual relevance of images to articles, global relevance, and potential for high impact on internet affordability given their significant traffic. The ranking was sourced from Similarweb Ltd., a leading software development and data aggregation company specializing in web traffic insights \cite{similarweb2024}. From each website, we randomly selected 5 categories (e.g., Sports, Business, Politics) based on the site's own categorization as defined in their navigation. For each category, we scraped the 10 most recent articles containing images, resulting in a total of 4,264 images across the 50 websites. This random selection of categories ensured a diverse range of topics, capturing both commonly featured and more unique categories across websites. Consequently, we gathered a representative dataset for evaluating image replaceability across various contexts, providing a robust foundation for assessing the potential of semantic caching in real-world scenarios.

For each article, we scraped all images in the article body\footnote{For 3 paywalled websites out of the 50 total, we were only able to extract the publicly accessible featured image of each article along with the article heading, both of which were available through the category listing pages.} along with their alternative text (alt text), where available, as well as the article heading. The alt text provided valuable contextual information, such as in the case of an article sourced from bbc.com; "Erica Fischer Mobile homes are highly combustible and parks are vulnerable during wildfires, as seen in this case during the 2018 Camp Fire in California." This kind of detail helps capture the specific circumstances, events, or even key figures depicted in an image, which is important for accurate image replacement decisions. 

\subsection{Ground Truth Labelling}
To create a dataset with ground truth labels, we had human annotators assign replaceability scores to every image pair in the dataset. The raters followed a systematic process: for each image pair, they opened the respective article and skimmed through it to gain a full understanding of the article's context before deciding if the two images could be swapped, and if so, at what cost to context loss to the user. 

Two independent sets of raters assessed replaceability across images from 3 websites each\footnote{One pair of raters evaluated 3058 image pairs, and the other rated a total of 4876 image pairs.}, and replaceability scores were assigned based on above mentioned approach. As a measure of inter-rater reliability, we used two-observer ordinal data version of the Krippendorff's Alpha~\cite{krippendorff_alpha} which yielded scores of 0.84 and 0.80, for the two pairs of raters respectively, indicating "strong" reliability~\cite{reliability_lvl}. In total, the labelled dataset consisted of 41,031 image pair evaluations across 50 websites.

Images in each category were evaluated for their replaceability with all other images in other articles of the same category. This comparison generated an $N \times N$ replaceability matrix, where $N$ represents the number of images in that category. To simplify comparisons, images from the same article were assigned a replaceability score of 0, as we are only considering inter-article image replacement. The ordinal replaceability scores range from 0 to 4, based on the following criteria: (i) 0: Not replaceable, (ii) 1: Somewhat replaceable, (iii) 2: Moderately replaceable, (iv) 3: Very replaceable, and (v) 4: Completely replaceable. 

We selected this scale to provide users or website providers with flexibility in opting for varying levels of image replaceability. Each score represents a different trade-off between minimizing context loss and maximizing data savings. A lower replaceability threshold (1 or 2) is suitable for the especially data constrained user as it would allow for more image replacements but at the cost of relatively significant contextual divergence. Conversely, higher thresholds (3 or 4) ensure minimal context loss but offer fewer opportunities for data reduction. 

\subsection{Evaluation Metrics}
\begin{figure}[h]
    \begin{minipage}[t]{0.25\textwidth}
        \centering
        \includegraphics[width=\linewidth]{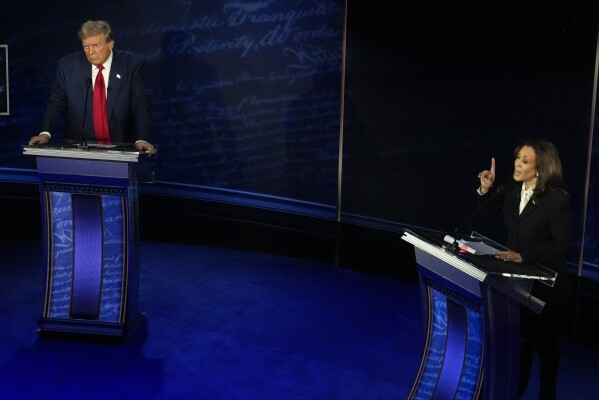}
        \subcaption[]{\footnotesize Base Image to replace}
    \end{minipage}
    \hfill
    \begin{minipage}[t]{0.48\textwidth}
        \centering
        \begin{minipage}[t]{0.48\textwidth}
            \centering
            \includegraphics[width=\linewidth]{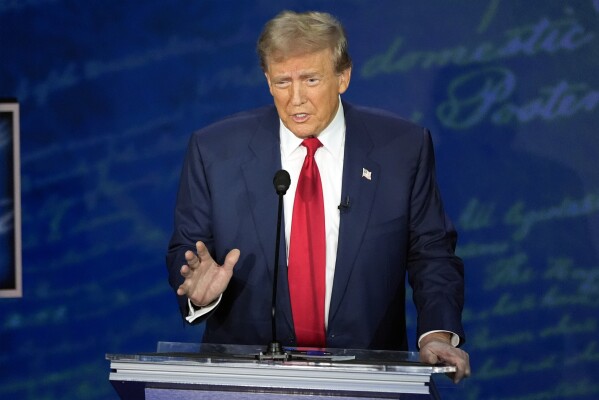}
            \subcaption[]{\footnotesize Image 1: Completely replaceable}
        \end{minipage}%
        \hfill
        \begin{minipage}[t]{0.48\textwidth}
            \centering
            \includegraphics[width=\linewidth]{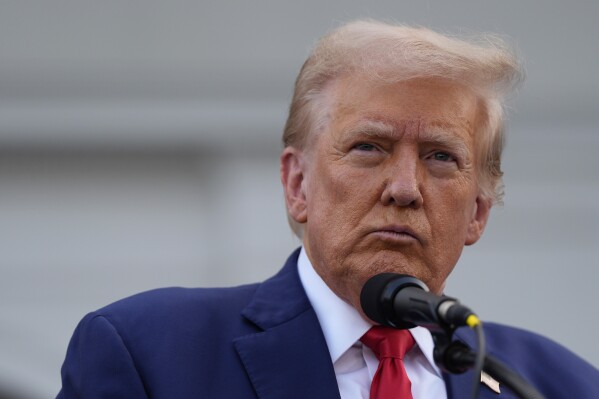}
            \subcaption[]{\footnotesize Image 2: Highly replaceable}
        \end{minipage}
        
        \vspace{6pt} 
        
        \begin{minipage}[t]{0.48\textwidth}
            \centering
            \includegraphics[width=\linewidth]{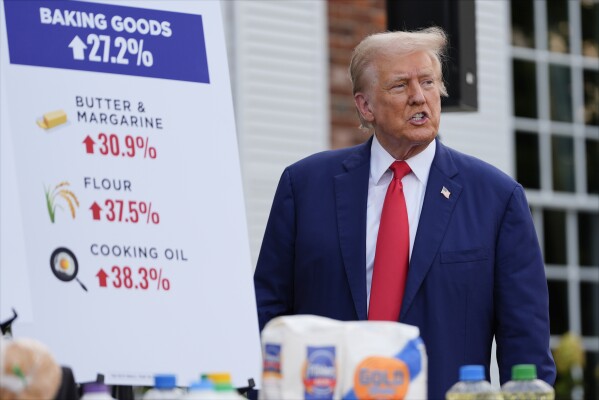}
            \subcaption[]{\footnotesize Image 3: Moderately replaceable}
        \end{minipage}%
        \hfill
        \begin{minipage}[t]{0.48\textwidth}
            \centering
            \includegraphics[width=\linewidth]{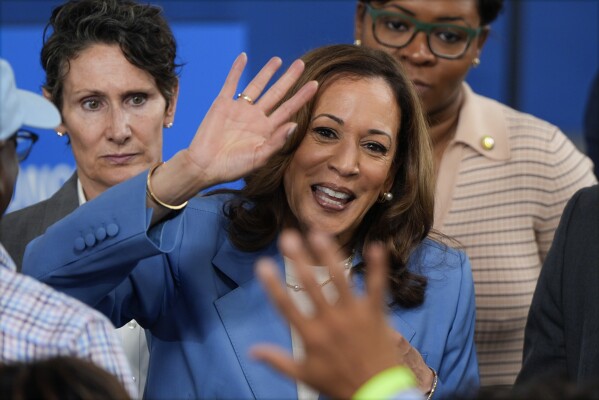}
            \subcaption[]{\footnotesize Image 4: Somewhat replaceable}
        \end{minipage}
    \end{minipage}
    \vspace{-0.15in}
\caption{\small{Human-annotated replaceability assessment for images 1-4 compared to the base image. Article headings and alt texts are included as contextual metadata.}} 
\label{image_examples}
\end{figure}

The following factors were considered by the human annotators while assigning replaceability scores:
\begin{itemize}[leftmargin=*,topsep=0pt]
\setlength{\itemsep}{0pt}
\setlength{\parskip}{0pt} 
    \item \textbf{Thematic Consistency}: Whether the images conveyed the same overarching theme and belonged to the same specific topic or niche (e.g., political rallies, a specific sporting event).
    \item \textbf{Relevance and Content Specificity}: How well each image aligned with the article’s context, such as depicting the main event or figure being discussed. Alternatively, if an image was more generic (e.g., an image of the United States Capitol), it could be considered easily replaceable across multiple articles within the same category.
    \item \textbf{Risk of Context Loss}: Consideration of the extent to which swapping the images would disrupt the audience's understanding of the article’s content.
    \item \textbf{Visual Style}: Whether the images shared similar visual elements, such as tone, mood, or composition, although this factor was considered secondary in comparison to thematic and contextual fit.
\end{itemize}

`Replaceability' is a semantically-driven metric that evaluates whether an image can substitute for another within the context of a different article from the same category. We assess this score based on whether the substitute image conveys similar thematic content or adds informational value to the article comparable to the value-add of the original image. We restrict comparisons to images within the same category because expanding the replaceability matrix to include images across all categories would result in a quadratic increase in the number of replaceability score computations— for each new image added to the dataset, N additional comparisons are required. However, images from different categories are unlikely to be contextually similar. By limiting comparisons to images within categories, we increase the likelihood of finding replaceable images while keeping the computational complexity manageable.

Based on these evaluation metrics, we illustrate examples of image pairs assigned replaceability rating on a scale of 0-4 in Figure \ref{image_examples}

In the examples above, the base image closely aligns with Image 1, as both depict the Trump-Harris presidential debate. The alt-text for both confirms this context, and the tone of a heated political exchange is consistent. Image 2, while not from the debate, shows Trump with a microphone, making it a good fit for any speech, with his tense expression matching the base image's mood. Image 3, though showing Trump speaking with similar tension, has a lower score as it captures a completely different event—a press conference in New Jersey, as indicated by the alt text. The elements in the image also clearly reference the press conference, unlike in image 3 which serves as a more generic image. Image 4, despite featuring Kamala Harris, a key figure, scores a 1 due to the significant shift in topic and tone, focusing on Harris's economic agenda, as reflected in the article heading.

%% file: potential.tex

\section{Potential For Semantic Web Caching}
\label{sec:potential}
Semantic caching maximises data reuse by leveraging the meaning and context behind image content, rather than relying on exact matches as in traditional caching mechanisms.

\subsection{Overview of Proposed Model}
Our system requires the server to store category information and an image ID (unique across a category) in the metadata of every image. By maintaining unique IDs only within a category instead of across the entire website, we reduce the encoding-space, thereby reducing the overhead incurred in client requests. Upon requesting for an image, the client should append the IDs of images in its cache from the same category, along with a threshold \( t \), an adjustable parameter between 1-4 that indicates the minimum replaceability required for image replacement.

On the server side, the similarity matrix for the relevant category is indexed to retrieve the similarity between the requested image \( I_{\text{req}} \) and the cached images \( I_{\text{cache}} \). If the maximum similarity score meets or exceeds the threshold \( t \), the server responds with the cached image \( I_{\text{cache}} \). If no cached image meets the threshold, the client downloads the requested image. This is accomplished through the following novel HTTP response directive:

\[
\verb|reuse_similar:| \arg\max_{I_{\text{cache}}} \left( \text{similarity}(I_{\text{req}}, I_{\text{cache}}) \geq t \right)
\]

\subsection{Potential for improving Internet Affordability}

\begin{figure}
  \centering
  \includegraphics[width=0.5\textwidth]{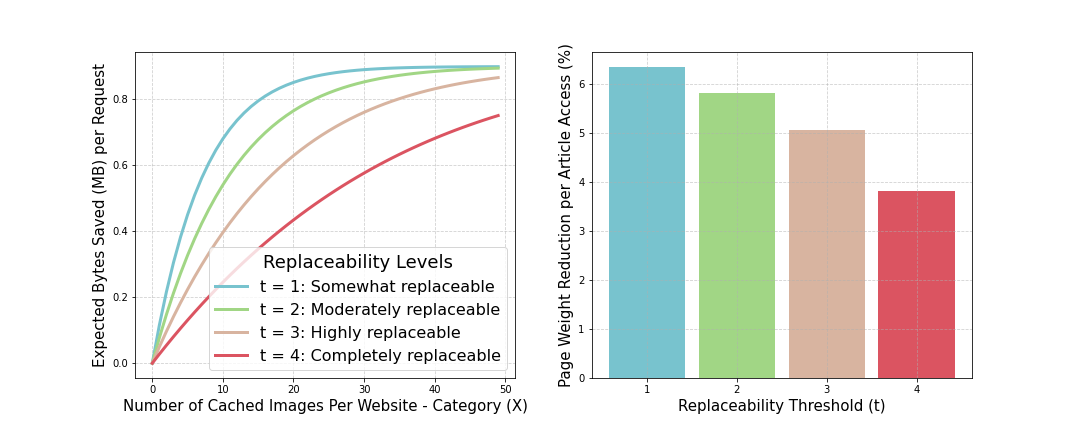}
  \caption{\small{Left: Expected byte savings as a function of relevant cached images. Based on HTTP Archive's median image size of 0.9 MB.
  Right: Reduction in data transfer as a percentage of page weight per article access, at ordinal levels of replaceability, using semantic caching. Based on our dataset of top 50 websites.}}
  \label{fig:expected_savings_per_req}
\end{figure}

To estimate the potential byte savings achieved through semantic caching, we developed a probability-based model to calculate the likelihood that an image request can be fulfilled from the cache. The main factors considered were the number of useful comparisons (i.e., images deemed semantically replaceable above a threshold), the size of the cache, and the byte size of the images.

Expected byte savings $\mu_t$, per HTTP request for an image can be modeled using the following equation:

\[
\mu_t = S.p - 2X
\]
where:
\[
p = 1 - \frac{\binom{N(1-u_t)}{X}}{\binom{N}{X}}
\]
\vspace{2mm}
\begin{itemize}
    \item $p$ is the probability that an image request is fulfilled from the cache.
    \item $u_t$ represents the percentage of useful comparisons above a similarity threshold $t$. For our dataset of news websites, we estimate $u_t \in [1.6\%, 9.5\%] \quad \text{for} \quad t = 1, \dots, 4$
    \item $N$ is the average number of comparisons per category. Across the 50 websites in our dataset, $N = 164$
    \item $S$ is the average image size which is estimated to be 0.9 MB based on the HTTP Archive\cite{httparchive2024} median value for image size (data collected from August-Sept 2024).
    \item $2X$ represents the overhead for appending $X$ image IDs, each encoded as a 16-bit (2-byte) hash, to the HTTP request. Here, $X$ is a subset of images within a category. A 16-bit hash provides a sufficient encoding-space under the assumption that image IDs are unique only within a category.
\end{itemize}

For our dataset of top 50 news websites, we computed the average image size \( S = 0.199 \, \text{MB} \), resulting in expected bytes saved per download request of \( 0.169 \, \text{MB} \) for \( t = 1 \) and \( 0.102 \, \text{MB} \) for \( t = 4 \). The average page weight was measured to be \( P = 4.77 \, \text{MB} \)\footnote{We use WebPageTest\cite{webpagetest} to measure the page weight from a sample of 1,000 websites in our dataset. This value is not taken from the HTTP Archive\cite{httparchive2024}, as we specifically focus on news websites, which tend to have more media content and are therefore expected to be heavier.}
, and the average number of images per article came out to \( I = 1.794 \).

Data savings  $\Mu_t$, per article access, as a fraction of page weight at similarity threshold $t$ can then be expressed as:

\[
\text{$\Mu_t$} = \frac{\mu_t \times I}{P}
\]

Hence, we estimate that on average, a user can achieve a reduction in total page weight of approximately \( 6.4\% \) for feasible replacements  (\( t = 1 \)) and \( 3.8\% \) for perfect replacements without incurring any context loss ( \( t = 4 \)).

Figure \ref{fig:expected_savings_per_req} illustrates the increase in byte savings per HTTP request with increase in cached images at client. Correspondingly, page weight reductions using semantic caching at the four levels of replaceability for our dataset of 50 websites is also shown. We observe that the variation in byte savings/request is relatively stable with standard deviation $\sigma \in [0.22, 0.24]$ MB across thresholds. This indicates moderate fluctuations in savings across different cache sizes while showing consistent trends in the efficacy of semantic reuse.

As expected, the rate of byte savings/request converges faster for lower thresholds. At \( t = 1 \), substantial savings are achieved with fewer cached images, and diminishing returns occur more quickly.

%% file: models.tex
\section{Leveraging LLMs for Semantic Web Caching}
\label{sec:llms}

Traditional visual similarity methods often fail to capture deeper, contextual relationships between images. Pre-trained convolutional neural networks (CNNs) like VGG16\cite{vgg16} or ResNet\cite{resnet} extract high-level features, allowing similarity computation through distance measures like cosine similarity or Euclidean distance between feature vectors. The Structural Similarity Index (SSIM)\cite{ssim} evaluates structural elements such as brightness and contrast. However, these approaches primarily focus on low-level visual features, struggling to capture semantic relationships. 

For tasks like image replacement, where understanding the image's meaning, context, and its value addition to the article is crucial, that too within the wider context of the article itself, incorporating semantics becomes necessary. This underscores the need for employing LLMs, which can better understand semantic meaning by processing text-based cues (e.g., alt text, headings) and make judgments based on themes, context, and tone.

\subsection{Model Evaluation Process}

In our evaluation, we utilized SoTA multi-modal LLMs, namely GPT-4o \cite{openai-gpt}, Gemini-1.5 pro \cite{gemini}, and Claude 3.5 Sonnet\cite{claude}, as well as the vision-language model LLaVA-NeXT \cite{li2024llava} as a precursor to preparing textual image descriptions input to LLaMA 3.1\cite{meta_LLaMA}. 
The two distinct approaches employed include:

\begin{itemize} [leftmargin=*,topsep=0pt]
\setlength{\itemsep}{0pt}
\setlength{\parskip}{0pt} 

\item \textbf{Direct Image-to-Output Pipeline:}
    In this approach, we directly input test image pairs and their associated article context into Claude, Gemini, and GPT-4o alongside the textual prompt. The models provided immediate output based on image comparisons and semantic understanding. 
    
    \item \textbf{Two-Step Image Description-to-Output Pipeline:}  
    
    Given the high costs associated with using proprietary models for extensive tasks like image replaceability judgments on large datasets, we also explored a more cost-effective solution. Here, we combined the open Large Multimodal Model (LMM) LLaVA-NeXT which is trained exclusively on image-text data, with LLaMA 3.1 for a two-step process: 
    
    \begin{enumerate}[leftmargin=*,topsep=0pt]
\setlength{\itemsep}{0pt}
\setlength{\parskip}{0pt} 
    \item LLaVA-NeXT was utilized to generate detailed image descriptions. LLaVA  was selected due to its superior image-text capabilities over reasoning,  with an accuracy of 90.92\% in generating image descriptions, rivalling the SoTA on this benchmark. \cite{liu2023llava}
    
    \item LLava generated descriptions, along with the article's contextual information were then input into LLaMA 3.1, an open-source model highly proficient in reasoning tasks, even outperforming GPT-4 in Abstraction and Reasoning Corpus (ARC) Challenge benchmark.\cite{meta_LLaMA} \\
    \end{enumerate}

    It is important to note that LLaVA-generated descriptions are valuable even in the presence of alt text because: (a) alt text may not always be available, and when it is, it often lacks sufficient context or detail; and (b) alt text is primarily used to identify specific figures or objects in the image which LLMs typically avoid for privacy reasons. While alt text provides specificity, LLaVA's more detailed and general descriptions offer a richer, broader understanding of the image content, complementing the alt text when present. \\

\end{itemize}

In contrast to the commercial LMMs, LLaVA-NeXT and LLaMA offer a highly scalable, open-source alternative. The compute and training costs for LLaVA-NeXT are approximately 100-1000 times smaller than other proprietary models\cite{liu2024llavanext}, and LLaMA is available freely, making this combination more accessible for large-scale applications. 

Each of the two approaches mentioned above were evaluated on zero-shot prompting as explained in the next section.

\subsection{Prompt Design}
\begin{figure}
  \centering
  \includegraphics[width=0.35\textwidth]{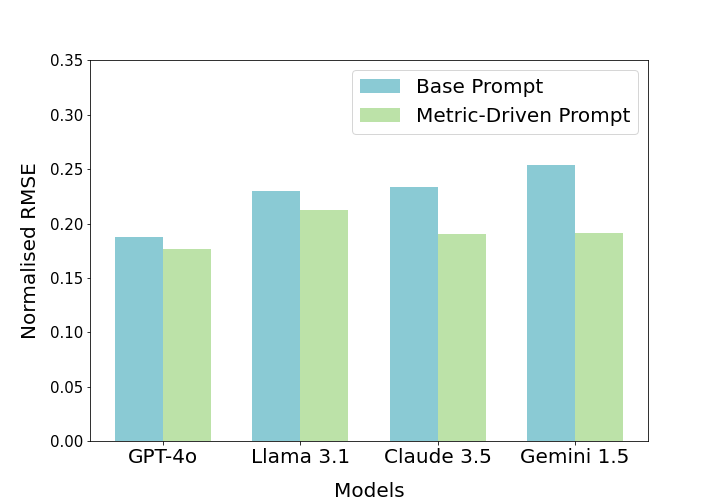}
  \vspace{-0.15in}
  \caption{\small{Normalized RMSE across four models using the base prompt with a general task description, compared to an evaluation metric-driven prompt generated by ChatGPT. All four models evaluated with zero-shot prompting}}
  \label{'prompt_eng'}
\end{figure}

Our prompt design process involved the development of two distinct prompts:

\begin{itemize}[leftmargin=*,topsep=0pt]
\setlength{\itemsep}{0pt}
\setlength{\parskip}{0pt} 
    \item \textbf{Base Prompt}: This initial prompt provided minimal guidance, allowing the LLM to independently assess semantic similarity. We did not provide any evaluation metrics and prompt the LLM to judge replaceability on our pre-defined scale of 0-4 given the test image pair (or image descriptions in the case of description-to-output pipeline)
    \item \textbf{Metric-Driven Prompt}: This version introduced specific evaluation metrics for the LLM to consider when assessing replaceability, offering brief but targeted direction to help the LLM make structured judgements.
\end{itemize}

Base Prompt was generated by the human evaluators while Metric-Driven Prompt was generated using ChatGPT-4 given instructions specifying the general metrics to consider as well as the flexibility to add further relevant factors.

Both prompts were developed using established best practices in prompt engineering, incorporating the following elements:
(i) Chain of Thought (CoT) reasoning: Implemented to encourage the model to provide step-by-step analytical responses. (ii) Role specification (``You are tasked with...''): Utilized to clearly define the LLM’s role and the objective of each task, enhancing task clarity and response accuracy, and (iii) XML tags: Employed for clear structural organization and to delineate components effectively.

The two prompts can be found in the \ref{appendix:prompts}

For the zero-shot setup, both prompts were tested across all models used in the two approaches on a sample of 5 websites\footnote{The selected websites include: www.aajtak.in, news.yahoo.com,  www.nypost.com, www.ndtv.com, and www.francetvinfo.fr} from the top 50 news sources in our dataset.

These websites were chosen to test the prompts on an overall dataset that has a relatively balanced distribution of all five labels used in our classification scheme.

We use normalized root mean square (NRMSE) as our primary evaluation metric because it effectively penalizes large deviations in ordinal data, and as a single statistic, has been shown to perform better than other measures~\cite{rmse}. While RMSE is traditionally applied to continuous data, its ability to handle small integer ordinal data makes it ideal for measuring the accuracy of our models' predictions on this task.

The results shown in Figure \ref{'prompt_eng'} reveal a consistent decrease in normalised NRMSE from the Base Prompt across all models, demonstrating the positive impact of more explicit guidance. For instance, Gemini-1.5 pro saw the largest RMSE reduction, lowering RMSE by $24.9\%$, indicating that the structured prompts helped reduce large prediction errors. Similarly, Claude 3.5 showed improvement, with its NRMSE falling by $18.4\%$ of its base prompt error.

We also explored in-context learning at inference time using dynamic few-shot prompting with LLaMA 3.1. However, the NRMSE increased by 0.032 compared to metric-driven zero-shot prompting. The detailed procedure and results are discussed in \ref{appendix:icl}.

%% file: analysis.tex
\section{Implementation and Results}
\label{sec:implementation}

\subsection{Proof of Concept Architecture}
\label{sec:proof_of_concept}
\begin{figure}
  \centering
  \includegraphics[width=0.35\textwidth]{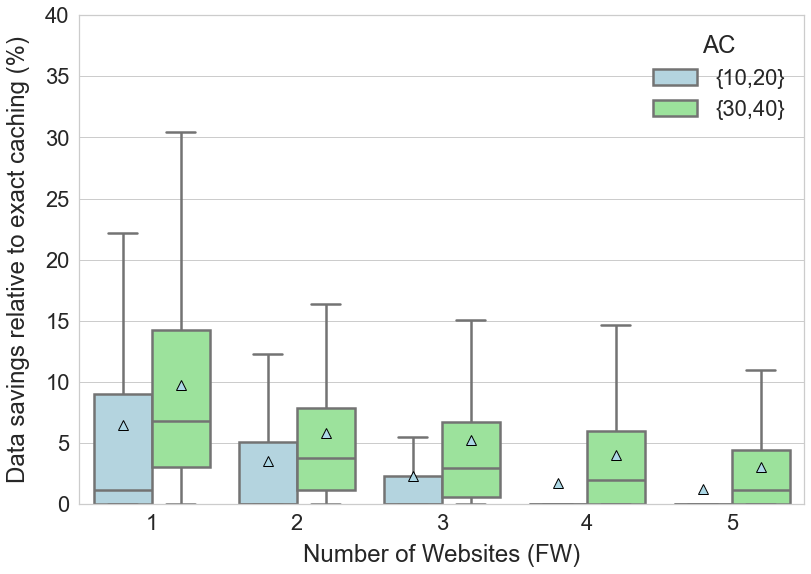}
  \caption{\small{Percentage reduction in data transfer compared to exact caching, with Frequented Websites (FW) ranging from 1 to 5. Total images requested (analogous to Articles Consumed (AC)) were 10, 20, 30, and 40, randomly sampled (with replacement) from FW. Each FW-AC combination was simulated with 100 random samples representing pseudo-clients. Replacement was triggered at the lowest threshold ("somewhat replaceable")}}
  \label{'boxplot'}
\end{figure}

We developed a client-server model to evaluate semantic caching's effectiveness in reducing data transfer for web images. Using our dataset of 4,264 images from 50 websites, we simulated random user requests to compare simple and semantic caching mechanisms.

On the server side, images were organized into categories, each linked to a replaceability matrix based on human annotations. In the semantic cache system, the client includes a list of relevant cached image IDs and a replaceability threshold $t$ in each image request, ensuring the IDs are from the same category and website as the requested image. The server then indexes into its replaceability matrix to determine if a cached image equals or exceeds the threshold $t$. If so, the server responds with \verb|reuse_similar =| $I_{cache}$, prompting the client to reuse the cached image and save bandwidth.

In this setup, image IDs are assumed to be unique within each category within a website.

To model user behavior, we focus on two variables: (a) Frequented Websites (FW), the number of news sites users often revisit, and (b) Articles Consumed (AC), the number of articles read, where requesting an article is analogous to fetching an image. We evaluate the model across various combinations of these variables to simulate a diverse set of browsing patterns.

From our dataset, FW websites are selected randomly (without replacement) 100 times to simulate user preferences. For each FW set, AC articles are selected randomly (with replacement), allowing for article revisits to capture exact cache savings. Each pseudo-client fetches a random set of images from their FW websites, first using a simple cache and then a semantic cache, recording the difference in data transfer.

Figure \ref{'boxplot'} illustrates data savings as FW and AC vary. Semantic caching shows significant data transfer reductions, averaging up to 9.8\% greater savings relative to exact caching. Savings are highest when users frequent fewer websites (FW = 1), with improvements over simple caching reaching up to $\approx 30\%$ for higher AC. As FW increases, the savings between AC groups decrease, but even with varied browsing (FW = 5, AC = 30, 40), average savings of around 3\% are observed, indicating semantic caching remains effective across different browsing habits.

\subsection{Overall Results}

\begin{figure}
  \centering
  \includegraphics[width=0.3\textwidth]{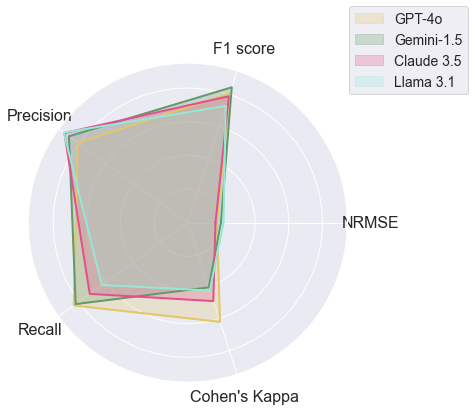}
  \caption{\small{Radar plot for  F1 score, precision, recall, and Cohen's kappa, weighted by class frequency, and normalised RMSE for the performance of each model on the task of assessing image replaceability. GPT-4o outperforms other models in predictive performance.}}
\label{radar}
\end{figure}

\begin{figure*}[ht]
  \centering
  \includegraphics[width=0.95\textwidth]{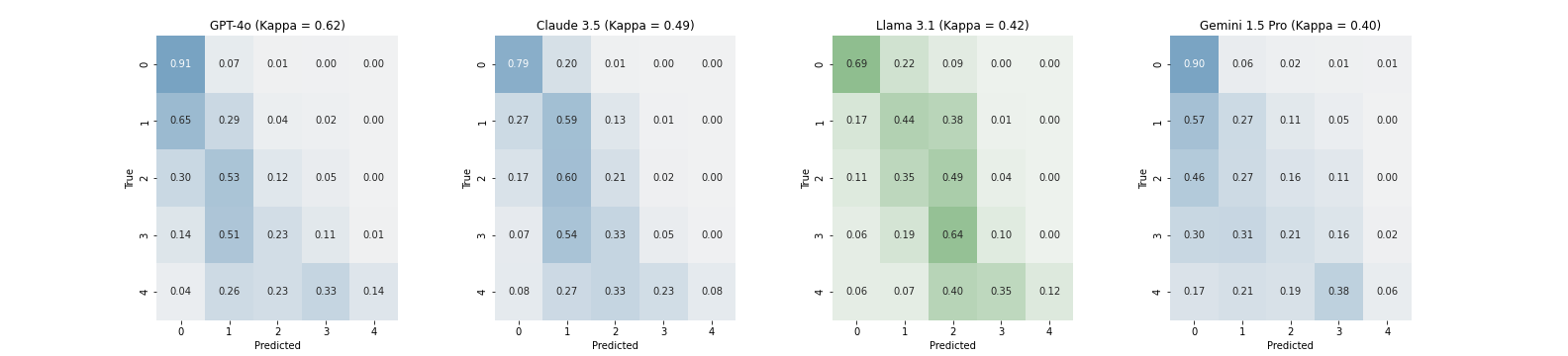}
  \caption{\small{Normalized (by row) confusion matrices for GPT-4o, Claude 3.5, LLaMA 3.1, and Gemini 1.5 Pro, illustrating the distribution of predicted vs. true labels across five classes. Darker colors along the diagonal indicate better classification accuracy, while off-diagonal shading reflects misclassification patterns.}}
  \label{fig:heatmap}
\end{figure*}

Using a dataset of 41,031 unique inter-article image pairs, we evaluated three commercial LLMs—GPT-4o\footnote{GPT-4o was tested on a subset of 20 websites due to cost constraints}, Gemini 1.5 Pro, and Claude 3.5 Sonnet—alongside the open-source LLaMA 3.1. Each model was tasked with rating the replaceability potential of image pairs on a scale of 0 to 4, based on the images themselves (or their descriptions in case of LLaMA 3.1), alt text, and article headlines using zero-shot prompting with COT reasoning.

To evaluate performance, we used the weighted F1 score, precision, and recall (accounting for class frequency), along with predictive accuracy measures such as normalized RMSE and weighted Cohen’s Kappa\cite{cohen1960coefficient}, a common metric for assessing agreement in ordinal categories.

The results are displayed in Figure \ref{radar} and detailed numerical results for each metric can be found in Appendix~\ref{appendix:results}. Notably, the open-source LLaMA 3.1, though more cost-effective, closely matched the performance of SOTA multi-modal models in several metrics, demonstrating the strong potential of open-source alternatives which could be more feasible to employ at a larger scale.

A closer look at the classification of image pairs is also provided in the confusion matrices in Figure \ref{fig:heatmap}, normalized by row to assess correct and incorrect predictions for each class. A common trend is underestimating replaceability, shown by the darker lower triangle in the heatmaps, where higher true values are often predicted as lower. This is especially pronounced in Gemini 1.5 Pro, where most predictions are 0. This explains why Gemini achieves the highest weighted F1 score (0.846), as class 0's dominance~\footnote{Approximately 90.5\% of image comparisons are human-labeled as 'not replaceable', with an average replaceability of $\mu = 12.38\%$ per category, $\sigma = 14.79\%.$} heavily influences the metric. However, this also results in Gemini having the highest  NRMSE and the lowest weighted Kappa, indicating only "fair" agreement\cite{kappa_categorization}.

LLaMA 3.1 performs best along the diagonal, especially for minority classes. However, its sensitivity to the dominant class 0 is the lowest (0.69), which reduces its overall Kappa score to 0.42, indicating "moderate agreement"\cite{kappa_categorization}, and increases its NRMSE to 0.21, the highest among the models. Claude 3.5 offers a more balanced prediction distribution, achieving the highest weighted precision (0.906), but its lower weighted recall (0.720) reduces accuracy for higher classes. Its Kappa score of 0.49 also reflects "moderate agreement." GPT-4o delivers the best performance, with the lowest NRMSE (0.173) and highest Kappa (0.62), indicating "substantial agreement"\cite{kappa_categorization} and the best overall accuracy. While it doesn't achieve the highest F1 score (0.807), it maintains a well-balanced performance across all classes.

The heatmaps' diagonal intensity alone doesn't fully reflect LLM performance, as the labels are ordinal. For Claude 3.5 and GPT-4o, adjacent scores carry significant weight, indicating frequent 1-point misclassifications due to the subjective nature of semantic judgments.

We also compute the variance in the LLM's assigned replaceability scores by repeatedly prompting with the same small set of test samples as detailed in ~\ref{appendix:results}.

\subsection{Cost Analysis}
We compared the estimated costs of running our pipelines on different popular LLMs. Specifically we evaluate the cost of comparing a pair of images for each LLM model (see the Appendix~\ref{appendix:cost-breakdown} for the detailed cost break down).

We recorded a cost of \$0 for LLaMA 3.1, as it is an open-source model that can be run locally, given adequate compute resources. This makes LLaMA 3.1 a unique option with no direct usage costs. Note that the other models are multi-modal, so we used the image-description-to-output pipeline discussed earlier only for LLaMA while the other models were evaluated and their cost was estimated on the direct image-to-output pipeline.

Among the commercial models, Gemini-1.5-pro stands out as the most affordable, with a cost of just \$0.0015 per comparison, while still delivering strong results when properly prompted. In comparison, GPT-4o costs \$0.00625, and Claude 3.5 Sonnet incurs the highest expense at \$0.0084 per comparison.

To assess cost-effectiveness, we can look at the ratio of performance (1/RMSE) to cost, excluding LLaMA due to its zero cost and slight difference in inputs. Based on this metric, Gemini offers the best performance-cost ratio, followed by GPT-4o, with Claude being the most expensive model in our analysis. Claude's higher cost may limit its feasibility for budget-conscious website operators. In contrast, Gemini presents a very good balance between performance and cost, making it a generally viable option.


%% file: discussion.tex
\section{Discussion and Limitations}
\textbf{Semantic Preservation.}
While our prompt design emphasized caution in swapping out images that feature specific events, key figures, or logos, the risk of replacing sensitive content remains with the use of LLMs. Generally, we have observed through the LLMs CoT explanations, that specificity in web content is especially penalised to reduce the likelihood of inappropriate swaps. However, certain edge cases still pose risks. For example, an article centered around a political figure might have their image mistakenly swapped with their political rival's if both names are present in the headline. Similarly, two visually similar but geographically different locations could be swapped, leading to misinformation, especially for data-constrained users relying on lower replaceability thresholds. 

\noindent \textbf{User-driven customization.} 
An important design choice for our system of caching is allowing the users to set a custom threshold based on their data usage preferences. The default threshold could be set to 0 (exact caching) in browser settings, but this would be adjustable by users themselves, offering them greater flexibility and control over the kind of context loss versus data savings trade-off they can expect.


\noindent \textbf{Incentives for website operators and potential resistance.} 
Website operators stand to gain from sending fewer bytes downstream, while potentially improving the performance of their pages for data-constrained users, leading to higher engagement. Moreover, operators retain the option to opt out of this system if they believe that the context loss from swapping images is too great. However, website operators may face certain integration challenges and costs in implementing semantic caching. They would need to generate unique image IDs across categories and embed the category and ID tags in the metadata of all images. Another challenge is the computational cost of evaluating replaceability for large image corpora. To address this, we demonstrate that our cost-effective LLaVA-LLaMA pipeline performs comparably to SOTA models across several metrics.

\noindent \textbf{Ethical Considerations.}
Ethical transparency is crucial to our caching system. Users should be aware that LLMs are being used to evaluate and swap images from their cache, and these images are visible to the website operators.

\noindent  \textbf{Limitations.}
In our measurement study, we focused solely on the ten most recent articles per category, operating under the assumption that users are more likely to engage with newer content. Expanding the time frame to include older articles would invariably increase the number of images available for comparison, which in turn would raise the number of semantic cache hits. It is worth investigating the increase in byte savings incurred by increasing the pool of articles. 

A limitation of this study is the restricted context provided to the LLMs for image replaceability assessment. We only supplied the article heading and alt text to limit input token length and conserve resources. A more comprehensive approach could involve adding the article summary or opening paragraph, which may improve accuracy but would require more computational resources. However, including the full article text may degrade performance. As prompts near the maximum token length, LLM reasoning tends to decline. For example, in the FLenQA dataset, accuracy drops from 0.92 for shorter inputs (around 250 tokens) to 0.68 for longer inputs (3000 tokens)\cite{moretokens}. This trend suggests that adding excessive context can lead to diminishing returns. Thus, the trade-off between additional context and model performance must be carefully considered to avoid significant accuracy losses.
\label{sec:discussion}

%% file: related_work.tex
\section{Related Work}
\label{sec:related_work}

In this section we explore past works that are related to our study. The relevant domains include improving web affordability, web and semantic caching, and use of LLMs for similar tasks.

\noindent \textbf{Web Affordability.} High data costs create significant barriers that prevent many underserved communities from accessing information on the internet. Several frameworks and approaches have been developed to address this challenge. The AW4A framework \cite{aw4a} focuses on webpage size reduction through modified web design principles, Google's WebLight \cite{weblight-deconst} optimizes web pages for low-bandwidth environments, Facebook's Free Basics \cite{fb-free-basics} provides free access to selected internet services. These solutions primarily target webpage complexity reduction \cite{rethinking-web-affordability} or create limited access models. In contrast, our work focuses on reducing web access costs through semantic image caching mechanisms, offering a novel approach that creates new opportunities for decreasing data consumption and improving affordability.

\noindent  \textbf{Web and Semantic Caching.}
Semantic caching represents an evolution of traditional caching, moving beyond exact matches to consider content meaning and context. Seminal work by Dar et al. \cite{semantic-caching-dar} demonstrated the value of maintaining semantic understanding in client-side caching, while Lee et al. \cite{Lee1999SemanticCV} explored semantic improvements to cache querying. These studies showed that incorporating semantic understanding could significantly increase cache hits while reducing cache misses. Our work advances this concept specifically for image caching in news websites. Previous research has explored feature-based similarity \cite{feature-based-similarity} in news contexts for recommender systems. Our approach leverages general-purpose LLMs to enable more flexible and contextually aware replaceability assessments rather than just article similarity.

\noindent  \textbf{LLMs for Semantic Caching.}
The use of semantic caching with LLMs is being explored actively. Recent systems like GPTCache \cite{bang-2023-gptcache} and SCALM \cite{SCALM} use semantic caching architectures to reduce costs associated with LLM usage. Our caching system is similar, but we apply it in the context of web objects on news websites, rather than for caching LLM queries and responses.

%% file: conclusion.tex
\section{Conclusion}
\label{sec:concl}
In this paper, we introduced a semantic caching system to improve web affordability by leveraging LLMs to evaluate inter-article image replaceability at various thresholds for data-constrained users. Our results show that semantic caching can significantly reduce data transfer, particularly for users who rely on a limited number of trusted news sources. We demonstrate that LLMs achieve reasonable accuracy in classifying replaceability, with the cost-effective solution of image description-based predictions by an open LLM, LLaMA 3.1 offering a viable alternative to the proprietary models. This approach highlights the potential of semantic caching for enhancing web access efficiency and reducing bandwidth consumption.

%% file: appendix.tex
\clearpage
\appendix 
\onecolumn
\section{Appendix}
\subsection{Prompts Used for Image Replaceability Evaluation}
\label{appendix:prompts}

\subsubsection{Base Prompt}
You are tasked with evaluating the replaceability of two images from different articles within the same category of a news website. 
Consider how well the two images align with each other in terms of their content and context. 

Use the following rating scale:

0: Not replaceable \\
1: Somewhat replaceable \\
2: Moderately replaceable \\
3: Very replaceable \\
4: Completely replaceable

Images and Associated Context:

\begin{verbatim}
<image_a>
    {{ image_a }}
</image_a>
<image_a_context>
    {{ image_a_context }}
</image_a_context>

<image_b>
    {{ image_b }}
</image_b>
<image_b_context>
    {{ image_b_context }}
</image_b_context> 
\end{verbatim} 
\vspace{1mm}
Using chain of thought prompting, analyze these two images and rate their replaceability. 
Break down your thought process step by step. 
Write your answer in the following format:

\begin{verbatim}
<rating>
[Your rating (0-4)]
</rating>

<justification>
Explanation: [Brief explanation for your rating, 
synthesizing your analysis of all factors]
</justification>
\end{verbatim}

\vspace{1cm}
\subsubsection{Metric-Driven Prompt}

You are tasked with evaluating the semantic replaceability of two images (Image A and Image B) from different articles within the same category of a news website. 
Your goal is to determine how interchangeable these images are based on their contexts and semantic similarity of the images, which include the article headings and alt text (where available).

Use the following rating scale for replaceability:
\begin{itemize}
    \item 0: Not replaceable
    \item 1: Somewhat replaceable
    \item 2: Moderately replaceable
    \item 3: Very replaceable
    \item 4: Completely replaceable
\end{itemize}

Here are the contexts for the two images:

\begin{verbatim}
<image_a>
    {{ image_a }}
</image_a>
<image_a_context>
    {{ image_a_context }}
</image_a_context>

<image_b>
    {{ image_b }}
</image_b>
<image_b_context>
    {{ image_b_context }}
</image_b_context>
\end{verbatim}
\vspace{1mm}
Consider the following factors when evaluating their semantic replaceability:
\begin{enumerate}
    \item Similarity of topics
    \item Specificity of information conveyed (e.g., specific people, places, etc.)
    \item Emotional tone or impact
    \item Potential for misinterpretation if swapped
\end{enumerate}

Using chain of thought prompting, analyze these two images and rate their replaceability. 
Break down your thought process step by step. 
Write your answer in the following format:

\begin{verbatim}
<rating>
[Your rating (0-4)]
</rating>

<justification>
Explanation: [Brief explanation for your rating, 
synthesizing your analysis of all factors]
</justification>
\end{verbatim}

\vspace{1cm}
\subsection{Label Counts Per Category}
\begin{table}[htbp]
    \caption{Counts of Labels for Image Pairs by Category. This table shows the distribution of labels assigned to image pairs across different categories. The non-zero label counts indicate the proportion of image pairs deemed replaceable.}
    \begin{tabular}{|p{0.2\linewidth}|p{0.06\linewidth}|p{0.06\linewidth}|p{0.06\linewidth}|p{0.04\linewidth}|p{0.04\linewidth}|p{0.12\linewidth}|p{0.12\linewidth}|}
        \hline
        \textbf{Category} & \textbf{0s} & \textbf{1s} & \textbf{2s} & \textbf{3s} & \textbf{4s} & \textbf{Non-Zero Labels} & \textbf{Total Labels} \\
        \hline
          \hline
        Health & 4102 & 117 & 117 & 51 & 38 & 323 & 4425 \\
        \hline
        Lifestyle & 2797 & 83 & 83 & 43 & 42 & 251 & 3048 \\
        \hline
        Science and Technology & 3795 & 114 & 79 & 25 & 32 & 250 & 4045 \\
        \hline
        World Affairs & 3659 & 84 & 49 & 44 & 57 & 234 & 3893 \\
        \hline
        Entertainment & 4108 & 125 & 89 & 62 & 46 & 322 & 4430 \\
        \hline
        Gender & 457 & 60 & 56 & 52 & 103 & 271 & 728 \\
        \hline
        Sports & 4641 & 132 & 189 & 120 & 121 & 562 & 5203 \\
        \hline
        Politics & 4952 & 130 & 135 & 94 & 97 & 456 & 5408 \\
        \hline
        Automotive & 1303 & 49 & 37 & 24 & 11 & 121 & 1424 \\
        \hline
    \end{tabular}
    \label{tab:label_counts_detailed}
\end{table}

\newpage
\subsection{Category Mappings}
\label{appendix:category-map}
\begin{table*}[ht]
    \caption{Mapping of General Categories to Specific Subcategories. We sampled 5 random categories per website in our dataset. To aggregate results by category, we use cosine similarity to map a specific subcategory from our original dataset to a generic category to evaluate extent of replaceability per general category}
    
    \begin{tabular}{p{0.15\linewidth} | p{0.75\linewidth}}
      \textbf{General Category}  & \textbf{Specific Categories} \\ \hline
      Business & Economy and Business, Mutual Funds, Economy, Travel, Relationship, Space, Executive, Business, Market Nightcap, Greater Jakarta Area, Crypto, Stock Market, Economics, Biden Economy, Finance, Discussion Club, Real Estate, Wine, Foreign, Mobile, Personal Finance, Books, Panorama, ISL, IT, Fintech World, oglobo-Business, UK, Showbiz \\
      \hline
      Health & Health, Medicare, Beauty, Food, Science \& Health, Health Sleep, Fitness, Health and Science, Food and Drink, oglobo-Cancer \\
      \hline
      Lifestyle & Lifestyle, Fashion, Gardening, Originals, Life \& Culture, Natural Wonders, Life, Climate, Life \& Environment, Environment, Society, oglobo-Brazil Environment, oglobo-Consumer Protection, Style, Style \& Beauty \\
      \hline
      Science and Tech & Technology, IT \& Science, Tech, Science Space, Science, Artificial Intelligence, Transport \& Energy, Climate Lab, Medscience, Universa, Space and Astronomy, Cybersecurity \\
      \hline
      World Affairs & U.S, World, Paris 2024, US \& Canada, Climate Solutions, World Middleeast Israel, International, Abroad, Israel War, Asia, British Royal Family, Latin America, Foreign Affairs, World News \\
      \hline
      Entertainment & Festivals, Music, Buzz, Education, Awards, Culture, Shows, Entertainment, oglobo-Entertainment, Cinema, Communication, Pop \\
      \hline
      Gender & Relationships, Gender, People, Domestic, Local, Abortion News, Prince Harry \\
      \hline
      Sports & Hockey, Electric, Animals, Racing, Golf, Sports Betting, Olympics, Tennis, Sports, Columns, Oscars, Football, Football Time Travel, Sport, Basketball, NFL, NBA, Weather, Cricket, Military \\
      \hline
      Politics & Politics, Crime, Immigration, 2024 Elections, India, War in Ukraine, Politics Congress, Police and Intelligence, US, News, Law, French Senate, Politics \& Economy, History, Policy, Follow the Candidates, Europe Politics, White House, US Elections, Russia Ukraine War, US Supreme Court, Fact Check, Elections, Psychology, Pakistan \\
      \hline
      Automotive & Auto, Car, Engines, Auto Racing, North Rhine-Westphalia, EPL, Automotive, Cars, Formula 1, Motorization, Aerospace \& Defense \\
    \end{tabular}
    \label{tab:categories}
\end{table*}

\newpage
\subsection{Detailed Results}
\label{appendix:results}

The following table provides the detailed numerical results for each model's performance in classifying replaceability of image pairs, complementing the radar plot in Figure \ref{radar}

\begin{table*}[ht]
    \caption{The models were tested on a total of 41,031 image pairs from the top 50 global news and media websites by traffic share\cite{similarweb2024}. Due to cost constraints, GPT-4o was evaluated on a representative sample of 20 websites from this set. Following are the performance metrics for GPT-4o, Gemini 1.5 Pro, Claude 3.5, and LLaMA 3.1 across various evaluation criteria. }

    \vspace{1cm}
    \begin{tabular}{lcccc}
        \toprule
        \textbf{Metric}           & \textbf{GPT-4o}   & \textbf{Gemini 1.5 Pro} & \textbf{Claude 3.5} & \textbf{LLaMA 3.1} \\
        \midrule
        \textbf{NRMSE}             & 0.17347           & 0.19688                & \textbf{0.16459}    & 0.21097            \\
        \textbf{F1 (weighted)}    & 0.80728           & \textbf{0.84586}       & 0.79002             & 0.72934            \\
        \textbf{Precision (weighted)} & 0.83760       & 0.87315                & 0.90649             & \textbf{0.91116}   \\
        \textbf{Recall (weighted)}    & 0.79695        & \textbf{0.82350}       & 0.72036             & 0.63062            \\
        \textbf{Response Variability ($s_p$)}          & 0.372             & 0                      & 0.2575              & 0.16               \\
        \bottomrule
    \end{tabular}
\end{table*}

The variability in model responses is quantified using the pooled standard deviation $s_p$\cite{pooled_std_wiki} across classes 0 to 4. For each class, a test pair that was accurately predicted by the model was re-prompted 20 times, totaling 100 runs across the 5 classes. This process captures the variation in the model's responses, indicating the consistency of the predictions. The temperatures for each model were kept at their default values: 1 for GPT-4o, Gemini 1.5 Pro, and Claude 3.5, and 0.8 for LLaMA 3.1.

\vspace{1cm}
\subsection{Cost Breakdown per Model}
\label{appendix:cost-breakdown}
\begin{table*}[ht]
    \caption{Cost-per-comparison estimates for Claude 3.5 Sonnet, GPT-4o, Gemini 1.5 Pro, and LLaMA 3.1. Input and output costs are calculated separately to give a total comparison cost.}
    \centering
    \begin{tabular}{lccc}
        \toprule
        \textbf{Model}           & \textbf{Input Cost Per Comparison (\$)} & \textbf{Output Cost Per Comparison (\$)} & \textbf{Total Cost Per Comparison (\$)} \\
        \midrule
        \textbf{Claude 3.5 Sonnet}  & 0.0039     & 0.0045     & 0.0084 \\
        \textbf{GPT-4o}             & 0.00325    & 0.003      & 0.00625 \\
        \textbf{Gemini 1.5 Pro}     & 0.00114    & 0.00038    & 0.0015 \\
        \textbf{LLaMA 3.1}          & 0*         & 0*         & 0* \\
        \bottomrule
    \end{tabular}
\end{table*}
Since input and output costs are calculated separately, we averaged the input and output tokens based on our dataset to estimate the total cost-per-comparison for each model. We took a conservative approach by overestimating the average input token length to 1300 tokens per image pair (including context) and an average output length of 300 tokens.

\twocolumn
\subsection{In-context Learning via Dynamic Few Shot prompting}
\label{appendix:icl}
\begin{figure}
  \centering
  \includegraphics[width=0.3\textwidth]{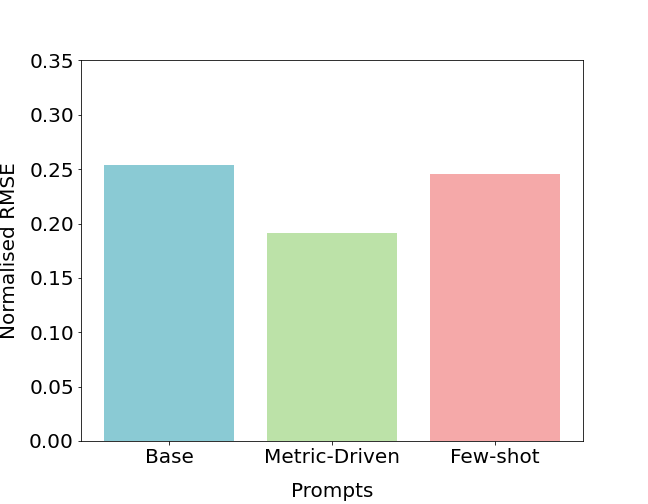}
  \caption{\small{Normalized RMSE for predictive accuracy of LLaMA 3.1 across zero-shot base and metric-driven prompting vs dynamic few-shot prompting by training with four relevant example pairs. NRMSE worsens with few-shot on this task.}}
  \label{'zero-few-shot'}
\end{figure}

In addition to zero-shot prompting, we explored in-context learning for the LLM by providing dynamic few-shot examples for the description-to-output pipeline\footnote{We evaluated only on LLaMA 3.1 across 29 websites from our dataset, excluding other commercial models due to cost constraints. As no performance improvement was observed for LLaMA, we did not extend this approach to the other models.} as shown in Figure \ref{'zero-few-shot'}

To create the few-shot dataset, we sampled image pairs with their metadata and true labels from the train set. Using cosine similarity, we first matched the test image pair's category with the most similar train category, and then identified the train image closest to test image A within that category. We selected five comparisons involving this train image as few-shot examples. We employ dynamic few-shot learning in an effort to condition the LLM at inference time by providing examples relevant to the test image pair.

Despite this, comparing against metric-driven zero-shot prompting, LLaMA's NRMSE increased by 0.032 with the inclusion of few-shot examples. We conjecture that this can be attributed to the limitations of its autoregressive architecture. Unlike bidirectional models, LLaMA processes information sequentially, which can cause problems in handling long input sequences that require distillation into brief outputs. In this case, the few-shot setup provided extensive context (image pairs and metadata) expecting a concise replaceability rating with brief explanation, likely causing confusion in the model's latent space~\cite{brown2020}. Autoregressive models often struggle with tasks demanding a complete understanding of large amounts of context or comparison of multiple pieces of information.